\documentclass[aps,twocolumn,nofootinbib]{revtex4}
\usepackage{bm,amssymb,epsf}
\newcommand{\kB}{k_{\rm B}}

\newcommand{\ave}[1]{\left\langle#1\right\rangle_\rho}

\newcommand{\cancor}[2]{\left\langle#1;#2\right\rangle_\rho}
\newcommand{\Qcommu}[2]{[#1,#2]}

\newcommand{\Qantico}[2]{\{#1,#2\}}
\newcommand{\CPoiss}[2]{\bm{\{}#1,#2\bm{\}}}
\newcommand{\Cdissip}[2]{\mbox{$\bm{[\hspace{-0.28em}[}$}#1,#2\mbox{$\bm{]\hspace{-0.28em}]}$}}

\begin{document}
\bibliographystyle{apsrev}

\title{The nonlinear thermodynamic quantum master equation}

\author{Hans Christian \"Ottinger}
\email[]{hco@mat.ethz.ch}
\homepage[]{http://www.polyphys.mat.ethz.ch/}
\affiliation{ETH Z\"urich, Department of Materials, Polymer Physics, HCI H 543,
CH-8093 Z\"urich, Switzerland}

\date{\today}

\begin{abstract}
The quantum master equation obtained by generalizing the geometric formulation of nonequilibrium thermodynamics to dissipative quantum systems is seriously nonlinear. We argue that nonlinearity occurs naturally in the step from reversible to irreversible equations and we analyze the nature of the nonlinear contribution. The thermodynamic nonlinearity leads to proper equilibrium solutions and improves the dynamic behavior of dissipative quantum systems. We also discuss the Markovian character of the thermodynamic quantum master equation. The general ideas are illustrated for two-level systems.
\end{abstract}

% \frac{\hbar}{2 \kB T_{\rm e}} at body temperature (310 K):  1.23 10^{-14} sec

% \pacs{05.70.Ln, 03.65.Yz}

% 03.65.Ta    Foundations of quantum mechanics; measurement theory
% 03.65.Yz    Decoherence; open systems; quantum statistical methods
% 05.10.Gg    Stochastic analysis methods (Fokker–Planck, Langevin, etc.)
% 05.20.-y    Classical statistical mechanics
% 05.30.-d    Quantum statistical mechanics
% 05.40.-a    Fluctuation phenomena, random processes, noise, and Brownian motion
% 05.60.-k    Transport processes
% 05.60.Cd    Classical transport
% 05.70.Ln    Nonequilibrium and irreversible thermodynamics
% 47.10.-g    General theory in fluid dynamics
% 47.45.-n    Rarefied gas dynamics
% 47.45.Ab    Kinetic theory of gases
% 51.10.+y    Kinetic and transport theory of gases

\maketitle

% ************
% Introduction
% ************

\emph{Introduction.}---For quantum information processing and a number of further present-day applications, dissipative quantum systems are a topic of key relevance. In a recent paper \cite{hco199}, thermodynamic arguments have been used to develop a master equation for quantum systems that are weakly coupled to a classical environment. This quantum master equation obtained from thermodynamics is significantly different from the master equations usually formulated with the sole guidance from principles of quantum mechanics \cite{Weiss,BreuerPetru}. For example, in the traditional approach, the linear Lindblad equation is obtained under a Markovian assumption \cite{Lindblad76}. The thermodynamic quantum master equation is neither linear nor Markovian. It is hence important to elaborate and illustrate some of its general properties, and that is the purpose of the present letter.

Whereas linearity seems natural in a quantum mechanical setting, it is quite unusual in thermodynamics. This is a consequence of the appearance of entropy, which typically involves logarithmic terms. Going beyond reversibility requires to go beyond linearity. As thermodynamics is the language for formulating healthy equations with well-behaved solutions, nonlinearity should not at all be considered as a drawback. Upon elaborating these general remarks in this letter, it should become increasingly clear that the thermodynamic quantum master equation is a most promising tool for all problems in quantum dissipation that have so far been treated by quantum master equations without a systematic and complete thermodynamic background. The thermodynamic quantum master equation allows us to couple quantum subsystems to all kinds of environments, including time-dependent ones, and it comes with a corresponding equation describing the influence of the quantum subsystem on the evolution of its environment.

% *************************************
% Thermodynamic quantum master equation
% *************************************

\emph{Thermodynamic quantum master equation.}---Based on thermodynamic considerations \cite{hco199}, the following master equation for the evolution of the density matrix or statistical operator $\rho$ on a Hilbert space ${\cal H}$ has been proposed to characterize a quantum subsystem in contact with an arbitrary classical nonequilibrium system acting as an environment:
\begin{eqnarray}
    \frac{d\rho}{dt} = \frac{i}{\hbar} \Qcommu{\rho}{H}
    &-& \frac{1}{\kB} \sum_j \Cdissip{H_{\rm e}}{S_{\rm e}}^j_x \, \Qcommu{Q_j}{\Qcommu{Q_j}{H}_\rho}
    \nonumber\\
    &-& \sum_j \Cdissip{H_{\rm e}}{H_{\rm e}}^j_x \, \Qcommu{Q_j}{\Qcommu{Q_j}{\rho}} .
    \qquad
\label{GENERICme}
\end{eqnarray}
In this equation, $\hbar$ and $\kB$ are Planck's constant (divided by $2\pi$) and Boltzmann's constant, respectively. The first term describes the reversible contribution to the evolution generated by the Hamiltonian $H$ via the commutator. All other terms are of irreversible nature and result from a coupling to the environment of the quantum subsystem. They are expressed through commutators involving the self-adjoint coupling operators $Q_j$ so that the normalization condition, ${\rm tr}\,\rho=1$, is automatically preserved in time. The subscript $\rho$ on a quantum observable indicates the modified self-adjoint operator
\begin{equation}\label{Atildef}
    A_\rho = \int_0^1  \rho^\lambda A \, \rho^{1-\lambda} \, d\lambda ,
\end{equation}
which is basically the product of $A$ and $\rho$, but with a compromise between placing $\rho$ to the left or the right of $A$.

Whereas the type of each coupling is given by the observable $Q_j$, the strength of a coupling is expressed in a dissipative bracket $\Cdissip{\,}{\,}^j$ defined as a binary operation on the space of observables for the classical environment (throughout this letter, boldface bracket symbols are used to distinguish classical brackets from quantum commutators and anticommutators). If the equilibrium or nonequilibrium states of the environment are characterized by state variables $x$, classical observables are functions or functionals of $x$, and their evaluation at a particular point of the state space is indicated by the subscript $x$. The classical observables $H_{\rm e}$ and $S_{\rm e}$ in Eq.~(\ref{GENERICme}) are the energy and the entropy of the environment, respectively. Dissipative brackets are bilinear, symmetric, $\Cdissip{A_{\rm e}}{B_{\rm e}}^j = \Cdissip{B_{\rm e}}{A_{\rm e}}^j$, positive, $\Cdissip{A_{\rm e}}{A_{\rm e}}^j \geq 0$, and satisfy the Leibniz or product rule, $\Cdissip{A_{\rm e} B_{\rm e}}{C_{\rm e}}^j = A_{\rm e} \Cdissip{B_{\rm e}}{C_{\rm e}}^j + B_{\rm e} \Cdissip{A_{\rm e}}{C_{\rm e}}^j$, for arbitrary environmental variables $A_{\rm e}$, $B_{\rm e}$, and $C_{\rm e}$.

% **************
% Basic features
% **************

\emph{Basic features.}---In view of the definition (\ref{Atildef}) of $A_\rho$, the second term in Eq.~(\ref{GENERICme}) will, in general, be nonlinear in $\rho$. This definition can be rewritten in a form that brings out the relationship to another possible compromise in placing $\rho$ and extracts the nonlinearity,
\begin{equation}\label{Atildefa}
    A_\rho = \frac{1}{2} \left( A \rho + \rho A + A'_\rho \right) , \quad
    A'_\rho =  \int_0^1 \Qcommu{\Qcommu{\rho^\lambda}{A}}{\rho^{1-\lambda}} \, d\lambda .
\end{equation}
Setting $A'_\rho = 0$ corresponds to the linearization of the master equation (\ref{GENERICme}) which has been proposed, but not recommended, in \cite{hco199}. We hence may think of $A'_\rho$ as the origin of nonlinearity in the full thermodynamically consistent quantum master equation. The importance of $A_\rho$ stems from its occurrence in the canonical correlation $\cancor{\,}{\,}$ (see Eq.~(4.1.12) of \cite{KuboetalII}),
\begin{equation}\label{cancor}
    \cancor{A}{B} = \int_0^1 {\rm tr}
    \big( \rho^\lambda A \, \rho^{1-\lambda} B \big) \, d\lambda
    = {\rm tr} \big( A_\rho B \big) ,
\end{equation}
which provides the key structural element in the geometric formulation of irreversible dynamics for quantum systems \cite{hco199}. The canonical correlation is symmetric, $\cancor{A}{B} = \cancor{B}{A}$, and positive, $\cancor{A}{A} \geq 0$. Moreover, averages can be obtained as special cases of canonical correlations, $\ave{A} = {\rm tr}(A \rho) = {\rm tr}(A_\rho) = \cancor{A}{1}$.

Whenever the state $x$ of the environment varies in time, the strength of the coupling in the thermodynamic quantum master equation (\ref{GENERICme}) becomes time-dependent. We then have a particular kind of non-Markovian behavior, as has been elaborated in the literature \cite{BreuerPetru,Maniscalcoetal04}. It is actually natural to expect that the environment changes in time because, in the thermodynamic approach \cite{hco199}, the quantum master equation (\ref{GENERICme}) comes together with a corresponding equation for the evolution of environmental observables,
\begin{eqnarray}
    \frac{dA_{{\rm e},x}}{dt} &=& \CPoiss{A_{\rm e}}{H_{\rm e}}_x
    + \Cdissip{A_{\rm e}}{S_{\rm e}}_x \nonumber\\
    &-& \frac{1}{\kB} \sum_j \Cdissip{A_{\rm e}}{S_{\rm e}}^j_x
    \cancor{\Qcommu{H}{Q_j}}{\Qcommu{H}{Q_j}} \nonumber\\
    &+& \sum_j \Cdissip{A_{\rm e}}{H_{\rm e}}^j_x \ave{\Qcommu{Q_j}{\Qcommu{Q_j}{H}}} .
\label{GENERICcla}
\end{eqnarray}
In this equation, $\CPoiss{\,}{\,}$ and $\Cdissip{\,}{\,}$ are the Poisson and dissipative brackets of the classical system, respectively \cite{hco99,hco100,hcobet}. In addition to the properties listed above for the dissipative brackets $\Cdissip{\,}{\,}^j$, energy conservation in the environment (except for exchange of energy with the quantum subsystem) is guaranteed by the degeneracy requirement $\Cdissip{A_{\rm e}}{H_{\rm e}} = 0$ for all observables $A_{\rm e}$. The Poisson bracket is bilinear, antisymmetric, and satisfies the Leibniz rule as well as the Jacobi identity, where the latter expresses the time-structure invariance of the Poisson bracket. All these properties of classical Poisson brackets are also satisfied by their quantum counterparts, the commutators.

If one looks only at the master equation (\ref{GENERICme}), the occurrence of time-dependent coefficients suggests non-Markovian behavior. If one looks at the coupled evolution for the quantum subsystem and the classical environment by the thermodynamically coupled set of Eqs.~(\ref{GENERICme}) and (\ref{GENERICcla}), however, the Markovian character of the description is restored, provided that the total system is closed. By comparing Eqs.~(\ref{GENERICme}) and (\ref{GENERICcla}) one notes obvious exchange terms between the two subsystems.

% ********************
% Equilibrium solution
% ********************

\emph{Equilibrium solution.}---The nonlinearity of the thermodynamic quantum master equation is essential for the existence of a proper equilibrium solution. If, for an environment in equilibrium with the temperature $T_{\rm e}$, we have the conditions
\begin{equation}\label{bathequilib}
    T_{\rm e} \Cdissip{H_{\rm e}}{S_{\rm e}}^j_{\rm eq} =
    \Cdissip{H_{\rm e}}{H_{\rm e}}^j_{\rm eq} ,
\end{equation}
for all $j$, then we obtain the proper equilibrium solution to Eq.~(\ref{GENERICme}),
\begin{equation}\label{rhoeq}
    \rho_{\rm eq} \, \propto \, \exp \left\{ - \frac{H}{\kB T_{\rm e}} \right\} .
\end{equation}
To obtain this result, we have made use of the identity $\Qcommu{A}{\rho} = \Qcommu{A_\rho}{\ln\rho}$ (see \cite{hco199}) in the last term of the master equation (\ref{GENERICme}) for $A=Q_j$.

The guaranteed existence of a proper equilibrium solution is a major advantage of the thermodynamic quantum master equation. It is deeply linked to the nonlinearity of the master equation.

% ****************
% Two-level system
% ****************

\emph{Two-level system.}---For a $k$-state (or, $k$-level) system, the underlying Hilbert space is a $k$-dimensional complex vector space which, without loss of generality, we can take as $\mathbb{C}^k$. The space of observables is the $k^2$-dimensional real vector space of self-adjoint $k \times k$-matrices with complex entries. For the two-level system, which has successfully been used to describe both nuclear magnetic resonance and spontaneous emission in quantum optics \cite{Hahn97}, we choose the $2 \times 2$-unit matrix $I$ and the three Pauli matrices
\begin{equation}\label{Pauli}
    \sigma_1 = \left(
      \begin{array}{rr}
        0 & 1 \\
        1 & 0 \\
      \end{array}
    \right) , \,\,
    \sigma_2 = \left(
      \begin{array}{rr}
        0 & -i \\
        i & 0 \\
      \end{array}
    \right) , \,\,
    \sigma_3 = \left(
      \begin{array}{rr}
        1 & 0 \\
        0 & -1 \\
      \end{array}
    \right) ,
\end{equation}
as the base vectors of the space of observables. More precisely, we express every self-adjoint complex $2 \times 2$-matrix $A$ in terms of a real scalar $\alpha$ and a real three-vector $\bm{a} = (a_1, a_2, a_3)$,
\begin{equation}\label{Arepresent}
    A = {\cal O}(\alpha, \bm{a}) = \frac{1}{2} \left( \alpha {\rm I}
    + a_1 \sigma_1 + a_2 \sigma_2 + a_3 \sigma_3 \right) .
\end{equation}
Note that $\alpha$ is the trace of $A$. Commutators and anticommutators can then conveniently be expressed as
\begin{equation}\label{commutator}
    \Qcommu{A}{B} = i {\cal O}(0,\bm{a}\times\bm{b}) ,
\end{equation}
\begin{equation}\label{acommutator}
    \Qantico{A}{B} = {\cal O}(\alpha\beta + \bm{a}\cdot\bm{b}, \beta\bm{a} + \alpha\bm{b} ) .
\end{equation}
From Eq.~(\ref{commutator}), we obtain an identity for the frequently occurring double commutators,
\begin{equation}\label{commutator2}
    \Qcommu{A}{\Qcommu{A}{B}} = {\cal O} (0, [a^2 \bm{1} - \bm{a}\bm{a}] \cdot \bm{b} ) ,
\end{equation}
where $\bm{1}$ is the $3 \times 3$-unit matrix. From Eq.~(\ref{acommutator}), we obtain
\begin{equation}\label{traceAB}
    2 \, {\rm tr}(AB) = \alpha\beta + \bm{a}\cdot\bm{b} .
\end{equation}
Arbitrary functions $f$ of an observable $A$ can be calculated with the formula
\begin{equation}\label{functionA1}
    f(A) = {\cal O} \left( f_+ + f_- , [f_+ - f_-] \, \bm{a}/a \right) ,
\end{equation}
with
\begin{equation}\label{functionA2}
    f_+ = f\left(\frac{\alpha+a}{2}\right) , \qquad
    f_- = f\left(\frac{\alpha-a}{2}\right) ,
\end{equation}
where $a=|\bm{a}|$. Equation (\ref{functionA1}) can be verified by induction for powers of $A$ and then generalized by Taylor expansion. From Eq.~(\ref{functionA1}) we further conclude that $(\alpha+a)/2$ and $(\alpha-a)/2$ must be the eigenvalues of $A$.

\begin{figure}
\centerline{\epsfxsize=5cm \epsffile{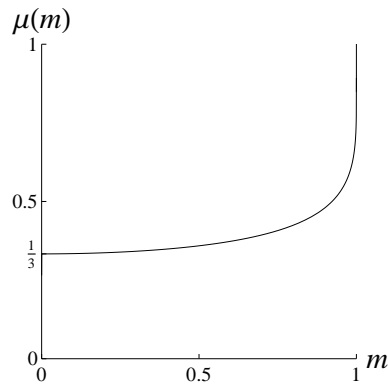}}
\caption[ ]{The function $\mu(m)$ characterizing the nonlinear contribution to the thermodynamic quantum master equation for a two-level system [see Eq.~(\ref{lambdawdef})].}
\label{fig_mu}
\end{figure}

As the density matrix has trace unity, it can be written as
\begin{equation}\label{densmatBloch}
    \rho = {\cal O}(1,\bm{m}) .
\end{equation}
For the eigenvalues to be nonnegative, we need $m = |\bm{m}| \leq 1$. This set of admissible choices of $\bm{m}$ is known as the Bloch sphere. For $m=1$, one of the two eigenvalues of $\rho$ is zero and we have a pure state. From Eq.~(\ref{traceAB}), we obtain $\ave{A} = (\alpha + \bm{a}\cdot\bm{m})/2$, which implies that the $j$th component of $\bm{m}$ is given by the average $\ave{\sigma_j}$. By using Eqs.~(\ref{commutator}) and (\ref{functionA1}) in Eq.~(\ref{Atildefa}), we find the following explicit form for the nonlinear part of $A_\rho$,
\begin{equation}\label{Atil2lev}
    A'_\rho = - {\cal O} \Big( 0, \mu(m)
    [ m^2 \, \bm{1} - \bm{m}\bm{m} ] \cdot \bm{a} \Big) ,
\end{equation}
with
\begin{equation}\label{lambdawdef}
    \mu(m) = \frac{1}{m^2} - \frac{1}{m \, {\rm artanh} \, m} .
\end{equation}
The function $\mu(m)$ is displayed in Figure~\ref{fig_mu}. The singularities of the two terms in Eq.~(\ref{lambdawdef}) at $m=0$ cancel so that $\mu(m) \approx 1/3$ for small $m$. According to Eq.~(\ref{commutator2}), the factor $[m^2 \, \bm{1} - \bm{m}\bm{m}]$ may be regarded as a double commutator formed with $\rho$. The nonlinear contribution to the quantum master equation associated with $\mu(m)$ leads to an improved relaxation behavior, as we shall see below.

We now choose the Hamiltonian $H = {\cal O}(0,\hbar \omega \bm{q}_3)$, where $\omega$ is the angular frequency associated with the energy difference between the two levels of the system and $\bm{q}_3 = (0,0,1)$, as well as the two coupling operators $Q_j =  {\cal O}(0,\bm{q}_j)$ with $\bm{q}_1 = (1,0,0)$ and $\bm{q}_2 = (0,1,0)$. The environment be a heat bath, the state of which be characterized by its energy $H_{\rm e}$. The thermodynamic properties of the bath are given by the thermodynamic relationship $S_{\rm e}(H_{\rm e})$, and we introduce the bath temperature by $1/T_{\rm e} = dS_{\rm e}/dH_{\rm e}$. This temperature characterizes the black-body radiation to which our system is exposed. Both dissipative brackets are assumed to be given by
\begin{equation}\label{CalLegdb}
    \Cdissip{A_{\rm e}}{B_{\rm e}}^j = \frac{dA_{\rm e}}{dH_{\rm e}}
    \, \gamma_0 \frac{\kB T_{\rm e}}{\hbar \omega} \, \frac{dB_{\rm e}}{dH_{\rm e}} ,
\end{equation}
where $\gamma_0$ is the spontaneous emission rate.

The quantum master equation (\ref{GENERICme}) can now be recognized to be equivalent to an evolution equation for $\bm{m}$, known as the Bloch equation \cite{Bloch46},
\begin{eqnarray}
    \frac{d\bm{m}}{dt} &=& \omega \, \bm{q}_3 \times \bm{m}
    - \gamma_0 \frac{2 \kB T_{\rm e}}{\hbar \omega} \bm{R} \cdot \bm{m}
    \nonumber \\
    &-& \gamma_0 \bm{q}_3
    + \gamma_0 \frac{\mu}{2} ( m^2 \, \bm{1} + \bm{m}\bm{m} ) \cdot \bm{q}_3 ,
\label{nlBlocheq}
\end{eqnarray}
with $\bm{R} = (\bm{1}+\bm{q}_3\bm{q}_3)/2$. Our choice of the coupling operators $Q_j$ is motivated by the two Lindblad operators that have been derived for quantum optical applications of two-level systems for the case of spontaneous emission (see, for example, Eq.~(3.219) of \cite{BreuerPetru}). As an alternative, one could include $Q_3 =  {\cal O}(0,\bm{q}_3)$ as a third operator with the same dissipative bracket (\ref{CalLegdb}) to achieve an isotropic frictional coupling. The only effect would be to change the anisotropic matrix $\bm{R}$ in Eq.~(\ref{nlBlocheq}) into the unit matrix $\bm{1}$. This would correspond to the case of strong collisions that cause energy decay whenever they cause dipole phase interruption \cite{BougouffaAlAwfi08}. For nuclear spin relaxation, for which the Bloch equation had been proposed originally, the situation with $\bm{R}=\bm{1}$ can also be realized, namely in isotropic molecular environments, both in gases and in low-viscosity liquids \cite{WangsnessBloch53}. It is well known, however, that longitudinal relaxation rates that are by orders of magnitude smaller than the transverse ones are much more typical for nuclear spin relaxation \cite{Bloch46}. This situation can be achieved by enhancing the coupling strength associated with $\bm{q}_3$ to become the dominating one.

The equilibrium solution of Eq.~(\ref{nlBlocheq}) is given by
\begin{equation}\label{eqsolm}
    \bm{m}_{\rm eq} = - \bm{q}_3 \, \tanh \left( \frac{\hbar \omega}{2 \kB T_{\rm e}} \right) .
\end{equation}
Contrary to the steady solution of the linearized quantum master equation, $\bm{m} = - \bm{q}_3 \hbar \omega/(2 \kB T_{\rm e})$, the solution (\ref{eqsolm}) always lies in the Bloch sphere, even for very low temperatures. The time-dependent solution of the thermodynamic quantum master equation can actually never leave the Bloch sphere, which is a nice consequence of thermodynamic consistency (note that, for very low temperatures, the solution of the linearized master equation leaves the Bloch sphere; although such low temperatures are unrealistic, this illustrates the improved relaxation behavior due to the nonlinearity). For a small deviation $\bm{m}'$ from the steady state solution (\ref{eqsolm}), we obtain a linearized version of Eq.~(\ref{nlBlocheq}),
\begin{eqnarray}
    \frac{d\bm{m}'}{dt} &=& \omega \, \bm{q}_3 \times \bm{m}'
    - \gamma_0 \frac{2 \kB T_{\rm e}}{\hbar \omega} \bm{R} \cdot \bm{m}'
    \nonumber \\
    &-& \gamma_0 \frac{m \mu}{2} (\bm{1}+3\bm{q}_3\bm{q}_3) \cdot \bm{m}'
    - \gamma_0 m^2 \frac{d\mu}{d m} \, \bm{q}_3\bm{q}_3 \cdot \bm{m}' , 
    \nonumber \\ &&
\label{nlBlocheqlin}
\end{eqnarray}
where $\mu(m)$ and its derivative are to be evaluated at $m_{\rm eq}$. The nonlinear terms enhance the relaxation, most dramatically near the boundary of the Bloch sphere.

The nonlinear part of $A_\rho$ in Eq.~(\ref{Atil2lev}) can be rewritten in the alternative form
\begin{equation}\label{Atil2levx}
    A'_\rho = \mu(m) \Big[ \Delta\rho \, {\rm tr}(A \Delta\rho) 
    - A \, {\rm tr}(\Delta\rho)^2 \Big] ,
\end{equation}
where $\Delta\rho = \rho - I/2$ is the deviation of the density matrix from a uniform distribution over all states (which corresponds to the entropy-dominated high-temperature limit). This reformulation suggests that the nonlinearity favors a uniform distribution and that the strength of this effect is proportional to $\mu(m)$. This is an effect of quantum fluctuations; it is independent of temperature and disappears for classical systems.

We had argued before that, in view of the presence of entropy, nonlinearity occurs very naturally in thermodynamics. How can the nonlinearity then disappear in the classical limit? This is indeed a 	 fortuitous cancelation occurring for classical systems on the level of distribution functions because of the identity $f d (\delta S/\delta f) = - \kB f d \ln f = - \kB d f$ (it is shown in Section III.B of \cite{hco99} that, in a thermodynamic approach, this property leads to the linearity of the Fokker-Planck equation for classical systems). The noncommutativity of quantum observables prevents such a cancelation.

% *******************
% Summary and outlook
% *******************

\emph{Summary and outlook.}---In this letter, we have established some advantages of the thermodynamically consistent quantum master equation. The resulting equation is significantly different from the ones commonly constructed without guidance from thermodynamics. In particular, the thermodynamic master equation is nonlinear and this nonlinearity is very helpful for obtaining physically meaningful solutions (for example, at equilibrium).

The thermodynamic quantum master equation describes the influence of any classical environment on a quantum subsystem. Moreover, it is supplemented by an equation describing the reverse influence of the quantum subsystem on the environment. If the total system is closed, we obtain a Markovian description of the coupled subsystems even if the coefficients in the quantum master equation change with a changing environment.

As quantum master equations are nowadays employed in many applications involving dissipative quantum systems, the nonlinear thermodynamic quantum master equation offers a new perspective on many problems. Problems that involve more complicated environments than simple heat baths can be approached in a thermodynamically consistent way. 

Our thermodynamic analysis is currently restricted to quantum master equations. Also for other approaches to dissipative quantum systems, such as operator Langevin equations, stochastic dynamics in Hilbert space, or path integrals \cite{Weiss,BreuerPetru,HanggiIngold05,FeynmanVernon63,Grabertetal88}, thermodynamic consistency should be established---this is work in progress.

% \vfill

% \bibliography{hcopubs}

\begin{thebibliography}{16}
\expandafter\ifx\csname natexlab\endcsname\relax\def\natexlab#1{#1}\fi
\expandafter\ifx\csname bibnamefont\endcsname\relax
  \def\bibnamefont#1{#1}\fi
\expandafter\ifx\csname bibfnamefont\endcsname\relax
  \def\bibfnamefont#1{#1}\fi
\expandafter\ifx\csname citenamefont\endcsname\relax
  \def\citenamefont#1{#1}\fi
\expandafter\ifx\csname url\endcsname\relax
  \def\url#1{\texttt{#1}}\fi
\expandafter\ifx\csname urlprefix\endcsname\relax\def\urlprefix{URL }\fi
\providecommand{\bibinfo}[2]{#2}
\providecommand{\eprint}[2][]{\url{#2}}

\bibitem[{\citenamefont{{\"O}ttinger}(2010)}]{hco199}
\bibinfo{author}{\bibfnamefont{H.~C.} \bibnamefont{{\"O}ttinger}},
The Geometry and Thermodynamics of Dissipative Quantum Systems,
arXiv:1002.2938v2 [quant-ph] 16 Feb 2010.

\bibitem[{\citenamefont{Weiss}(2008)}]{Weiss}
\bibinfo{author}{\bibfnamefont{U.}~\bibnamefont{Weiss}},
  \emph{\bibinfo{title}{Quantum Dissipative Systems}}, Series in Modern
  Condensed Matter Physics, Volume~13 (\bibinfo{publisher}{World Scientific},
  \bibinfo{address}{Singapore}, \bibinfo{year}{2008}), \bibinfo{edition}{3rd}
  ed.

\bibitem[{\citenamefont{Breuer and Petruccione}(2002)}]{BreuerPetru}
\bibinfo{author}{\bibfnamefont{H.-P.} \bibnamefont{Breuer}} \bibnamefont{and}
  \bibinfo{author}{\bibfnamefont{F.}~\bibnamefont{Petruccione}},
  \emph{\bibinfo{title}{The Theory of Open Quantum Systems}}
  (\bibinfo{publisher}{Oxford University Press}, \bibinfo{address}{Oxford},
  \bibinfo{year}{2002}).

\bibitem[{\citenamefont{Lindblad}(1976)}]{Lindblad76}
\bibinfo{author}{\bibfnamefont{G.}~\bibnamefont{Lindblad}},
  \bibinfo{journal}{Commun.\ Math.\ Phys.} \textbf{\bibinfo{volume}{48}},
  \bibinfo{pages}{119} (\bibinfo{year}{1976}).

\bibitem[{\citenamefont{Kubo et~al.}(1991)\citenamefont{Kubo, Toda, and
  Hashitsume}}]{KuboetalII}
\bibinfo{author}{\bibfnamefont{R.}~\bibnamefont{Kubo}},
  \bibinfo{author}{\bibfnamefont{M.}~\bibnamefont{Toda}}, \bibnamefont{and}
  \bibinfo{author}{\bibfnamefont{N.}~\bibnamefont{Hashitsume}},
  \emph{\bibinfo{title}{Nonequilibrium Statistical Mechanics}},
  vol.~\bibinfo{volume}{II} of \emph{\bibinfo{series}{Statistical Physics}}
  (\bibinfo{publisher}{Springer}, \bibinfo{address}{Berlin},
  \bibinfo{year}{1991}), \bibinfo{edition}{2nd} ed.

\bibitem[{\citenamefont{Maniscalco et~al.}(2004)\citenamefont{Maniscalco,
  Intravaia, Piilo, and Messina}}]{Maniscalcoetal04}
\bibinfo{author}{\bibfnamefont{S.}~\bibnamefont{Maniscalco}},
  \bibinfo{author}{\bibfnamefont{F.}~\bibnamefont{Intravaia}},
  \bibinfo{author}{\bibfnamefont{J.}~\bibnamefont{Piilo}}, \bibnamefont{and}
  \bibinfo{author}{\bibfnamefont{A.}~\bibnamefont{Messina}},
  \bibinfo{journal}{J.~Opt.~B: Quantum Semiclass.\ Opt.}
  \textbf{\bibinfo{volume}{6}}, \bibinfo{pages}{S98} (\bibinfo{year}{2004}).

\bibitem[{\citenamefont{Grmela and {\"O}ttinger}(1997)}]{hco99}
\bibinfo{author}{\bibfnamefont{M.}~\bibnamefont{Grmela}} \bibnamefont{and}
  \bibinfo{author}{\bibfnamefont{H.~C.} \bibnamefont{{\"O}ttinger}},
  \bibinfo{journal}{Phys.\ Rev.\ E} \textbf{\bibinfo{volume}{56}},
  \bibinfo{pages}{6620} (\bibinfo{year}{1997}).

\bibitem[{\citenamefont{{\"O}ttinger and Grmela}(1997)}]{hco100}
\bibinfo{author}{\bibfnamefont{H.~C.} \bibnamefont{{\"O}ttinger}}
  \bibnamefont{and} \bibinfo{author}{\bibfnamefont{M.}~\bibnamefont{Grmela}},
  \bibinfo{journal}{Phys.\ Rev.\ E} \textbf{\bibinfo{volume}{56}},
  \bibinfo{pages}{6633} (\bibinfo{year}{1997}).

\bibitem[{\citenamefont{{\"O}ttinger}(2005)}]{hcobet}
\bibinfo{author}{\bibfnamefont{H.~C.} \bibnamefont{{\"O}ttinger}},
  \emph{\bibinfo{title}{Beyond Equilibrium Thermodynamics}}
  (\bibinfo{publisher}{Wiley}, \bibinfo{address}{Hoboken},
  \bibinfo{year}{2005}).

\bibitem[{\citenamefont{Hahn}(1997)}]{Hahn97}
\bibinfo{author}{\bibfnamefont{E.~L.} \bibnamefont{Hahn}},
  \bibinfo{journal}{Concepts Magn.\ Reson.} \textbf{\bibinfo{volume}{9}},
  \bibinfo{pages}{69} (\bibinfo{year}{1997}).

\bibitem[{\citenamefont{Bloch}(1946)}]{Bloch46}
\bibinfo{author}{\bibfnamefont{F.}~\bibnamefont{Bloch}},
  \bibinfo{journal}{Phys.\ Rev.} \textbf{\bibinfo{volume}{70}},
  \bibinfo{pages}{460} (\bibinfo{year}{1946}).

\bibitem[{\citenamefont{Bougouffa and Al-Awfi}(2008)}]{BougouffaAlAwfi08}
\bibinfo{author}{\bibfnamefont{S.}~\bibnamefont{Bougouffa}} \bibnamefont{and}
  \bibinfo{author}{\bibfnamefont{S.}~\bibnamefont{Al-Awfi}},
  \bibinfo{journal}{J.~Mod.\ Opt.} \textbf{\bibinfo{volume}{55}},
  \bibinfo{pages}{473} (\bibinfo{year}{2008}).

\bibitem[{\citenamefont{Wangsness and Bloch}(1953)}]{WangsnessBloch53}
\bibinfo{author}{\bibfnamefont{R.~K.} \bibnamefont{Wangsness}}
  \bibnamefont{and} \bibinfo{author}{\bibfnamefont{F.}~\bibnamefont{Bloch}},
  \bibinfo{journal}{Phys.\ Rev.} \textbf{\bibinfo{volume}{89}},
  \bibinfo{pages}{728} (\bibinfo{year}{1953}).

\bibitem[{\citenamefont{H{\"a}nggi and Ingold}(2005)}]{HanggiIngold05}
\bibinfo{author}{\bibfnamefont{P.}~\bibnamefont{H{\"a}nggi}} \bibnamefont{and}
  \bibinfo{author}{\bibfnamefont{G.-L.} \bibnamefont{Ingold}},
  \bibinfo{journal}{Chaos} \textbf{\bibinfo{volume}{15}},
  \bibinfo{pages}{026105} (\bibinfo{year}{2005}).

\bibitem[{\citenamefont{Feynman and Vernon}(1963)}]{FeynmanVernon63}
\bibinfo{author}{\bibfnamefont{R.~P.} \bibnamefont{Feynman}} \bibnamefont{and}
  \bibinfo{author}{\bibfnamefont{F.~L.} \bibnamefont{Vernon}},
  \bibinfo{journal}{Ann.\ Phys.\ (N.Y.)} \textbf{\bibinfo{volume}{24}},
  \bibinfo{pages}{118} (\bibinfo{year}{1963}).

\bibitem[{\citenamefont{Grabert et~al.}(1988)\citenamefont{Grabert, Schramm,
  and Ingold}}]{Grabertetal88}
\bibinfo{author}{\bibfnamefont{H.}~\bibnamefont{Grabert}},
  \bibinfo{author}{\bibfnamefont{P.}~\bibnamefont{Schramm}}, \bibnamefont{and}
  \bibinfo{author}{\bibfnamefont{G.-L.} \bibnamefont{Ingold}},
  \bibinfo{journal}{Phys.\ Rep.} \textbf{\bibinfo{volume}{168}},
  \bibinfo{pages}{115} (\bibinfo{year}{1988}).

\end{thebibliography}

\end{document}